\documentclass{cs20proc}

\usepackage{kantlipsum}

\editors{S. J. Wolk}
\publisher{Zenodo}
\conference{The 20th Cambridge Workshop on Cool Stars, Stellar Systems, and the Sun}
\conferencedate{2018}

\title{Rotation and magnetic activity evolution of Sun-like stars}
\author{Philippe Gondoin
        }

\affiliation{European Space Agency, ESTEC - Postbus 299, 2200 AG Noordwijk, The Netherlands}

\shorttitle{Magnetic activity on Sun-like stars}
\shortauthors{P. Gondoin}

\abs{Characterising the long-term evolution of magnetic activity on Sun-like stars is important not only for stellar physics but also for understanding the environment in which planets evolve.  In the past decades, many photometric surveys of open clusters have produced extensive rotation period measurements on Sun-like stars of different ages. I use these data to infer the long-term evolution of Ca II chromospheric and X-ray coronal emission on solar mass stars combining a best fit parametric model of their rotation evolution with empirical rotation-activity relationships.
}

\begin{document}

\maketitle

\section{Introduction}

Characterising the evolution of magnetic activity on Sun-like stars is important not only for stellar physics, but also for understanding the environment in which planets evolve. Magnetic activity can have a strong impact on the long-term evolution of planetary atmospheres (e.g. Kulikov et al. 2007; Lammer et al. 2008; Forget \& Leconte 2014).  Magnetic activity signatures include the emission reversal in the core of the Ca II H\,\&\,K lines (see Hall 2008 and reference therein) and the coronal X-ray emission (e.g. Schrijver \& Zwaan 2000; Petsov et al. 2003).

Skumanich (1972) initially reported that a comparison of the Ca II emission luminosity for the Pleiades, Ursa Major, and Hyades stars and the Sun indicates an emission decay which varies as the inverse square root of the age.  More recently,  Pace \& Pasquini (2004) argue that the decay of the chromospheric activity between 500 Myrs and 2 Gyrs is similar but steeper than the inverse square root of time. Mamajek and Hillenbrand (2008) derive an improved activity-age calibration for solar-type dwarfs between the age of the Hyades and that of the Sun by combining cluster activity data with modern cluster age estimates.

Regarding the coronal X-ray emission, the current picture (see G\"udel 2007 and references therein) is that solar mass stars enter the main sequence in a saturation regime with little evolution of their X-ray and UV luminosity. At ages up to 200 Myr, the stars coronae drop out of the saturation regime and their X-ray luminosity decays as t$^{-1.5}$ where t is the stellar age (Maggio et al. 1987; G\"udel et al. 1997). 

These relationships between chromospheric or coronal activity indicators and ages provide a simple description of the long-term evolution of magnetic activity on Sun-like stars. However, they do not account for the wide spread of activity levels that are observed in young clusters among solar mass stars of similar ages. Neither do they take into account the rotation history of individual stars that play a role in the evolution of their activity level (Gondoin 2012, 2013). 

The dependence on stellar mass, age and rotation renders a direct measurement of the magnetic activity evolution difficult. In addition, the long-term evolutionary trend of activity indicators is blurred by the short-time variability of magnetic phenomena that includes the evolution of active regions and activity cycles. The determination of an average magnetic activity level at different ages thus requires observations of large numbers of coeval stars.  Such samples are found in open clusters but their observations are limited by the low brightnesses of activity indicators in weakly or moderately active stars and by the large distances of clusters.  An alternative approach is to measure activity indices on a large number of field stars but their ages are seldom known to sufficient accuracy. 

In order to overcome these difficulties, I characterise the long-term evolution of magnetic activity on Sun-like stars indirectly. The approach consists of combining a best fit parametric model of their rotation evolution with rotation-activity relationships.  Section 2 describes the derivation of the best fit parameters of a rotation evolution model using measurements of stellar rotation periods in open clusters of various ages. Section 3 combines this best fit model with rotation-activity relationships to calculate the long-term evolution of chromospheric and coronal activity indices. Section 4 compares the derived description of the magnetic activity evolution on Sun-like stars with measurements of activity indicators in open clusters. The results are summarised in Sect. 5.

\begin{figure*}[!ht]
\begin{tabular}{l l}
	\centering
 \includegraphics[width=0.48\linewidth, angle=0]{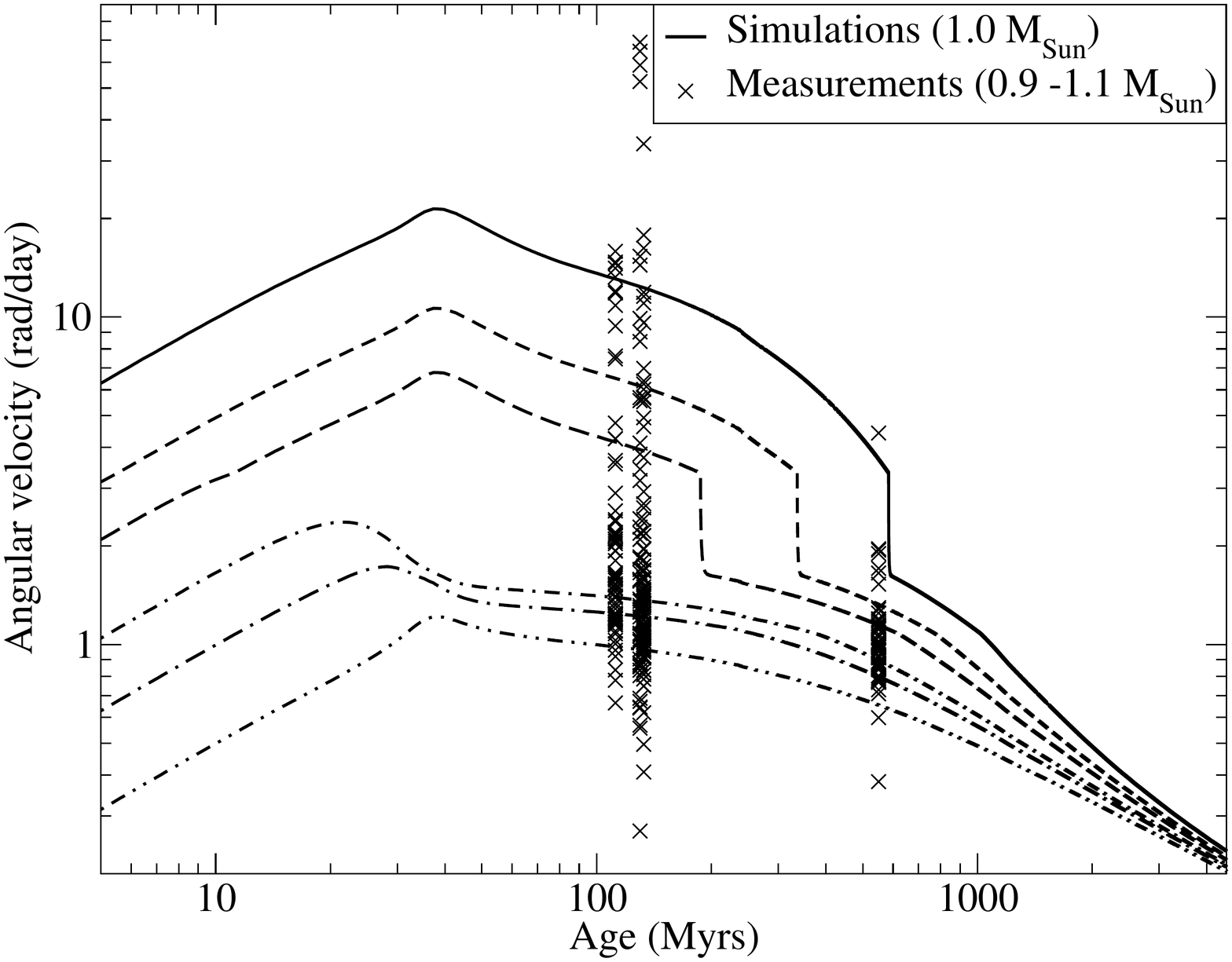} & \includegraphics[width=0.48\linewidth, angle=0]{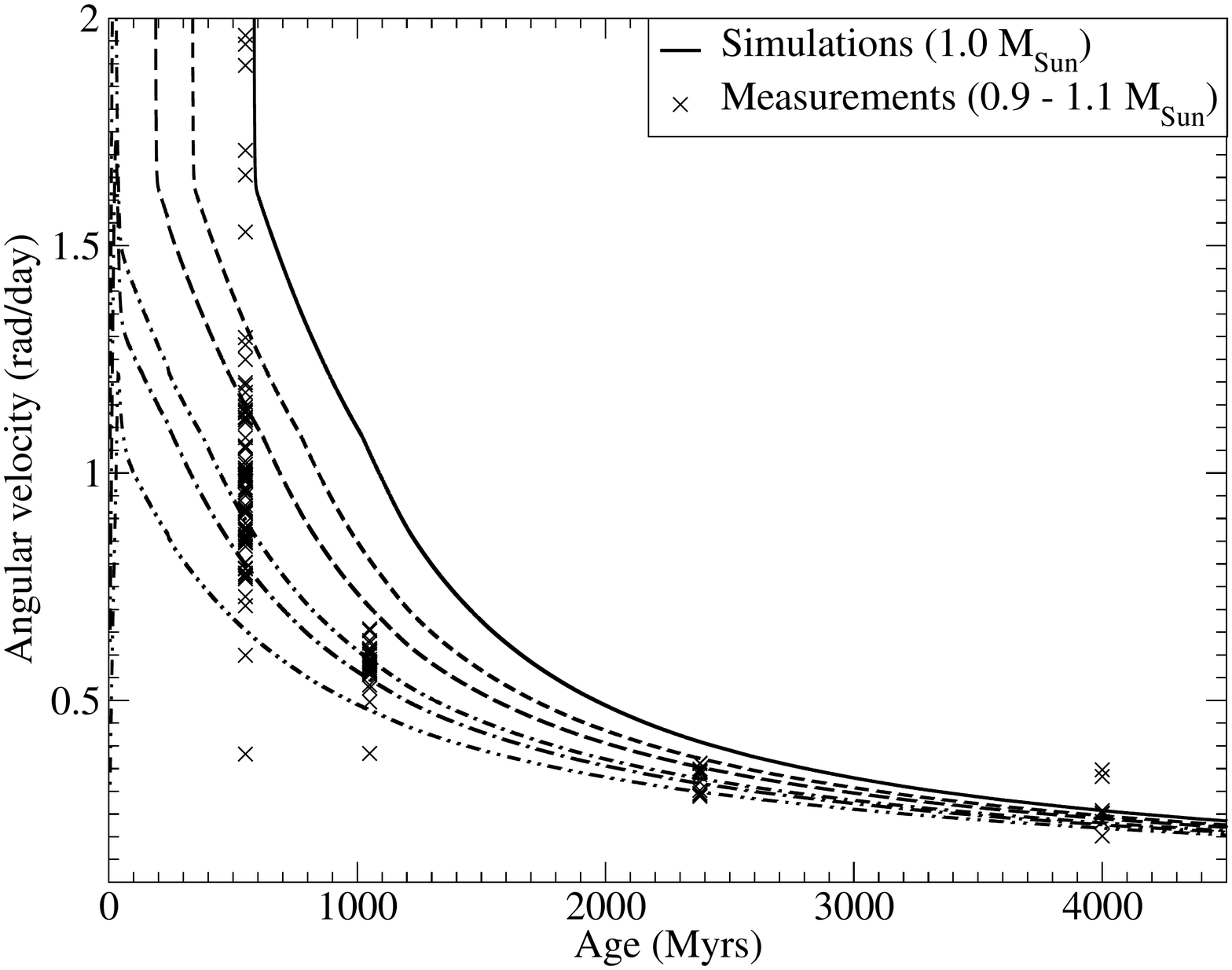} \\ 	 	
\end{tabular}
	\caption{Simulated angular velocity evolutions of 1.0 M$_{\rm \odot}$ stars with initial rotation periods at 5 Myrs of 1 (upper curve), 2, 3, 6, 10, and 20 days (lower curve). Left: the simulated angular velocity curves in log-log scale are compared with rotation measurements of 0.9 - 1.1 M$_{\rm \odot}$ stars (black crosses) in the Pleiades (112 Myrs; Hartman et al. 2010), M50 (130 Myrs; Irwin et al. 2009), M35 (133 Myrs; Meibom et al. 2009), and M37 (550 Myrs; Hartman et al. 2009).  Right: the simulated angular velocity curves in linear scale are compared with rotation measurements of 0.9 - 1.1 M$_{\rm \odot}$ stars (black crosses) in M37, NGC 6811 (1.05 Gyrs; Meibom et al. 2011), NGC 6819 (2.38 Gyrs; Meibom et al. 2015), and M67 (4.0 Gyrs; Barnes et al. 2016).}
	\label{fig:fig_meas_rot_distrib}
\end{figure*}

\section{The rotation evolution of Sun-like stars}

Gondoin (2016, 2017) describes a parametric model of rotation evolution on Sun-like stars obtained by combining the so-called double zone model (see e.g MacGregor \& Brenner 1991; Keppens et al. 1995; Allain 1998; Spada et al. 2011; Oglethorpe \& Garaud 2013) with the one dimensional model of Weber \& Davis (1967) that quantifies the torque exerted by the wind on the star. The combined model model has four free parameters, namely (i) the initial angular rotation velocity of the star after dispersion of its circumstellar disk, (ii) the mass loss rate of the stellar wind at fast rotation, (iii) the Alfven radius at a given rotation rate or Rossby number, and (iv) the coupling timescale between the radiative core and the convective envelope of the star that are assumed to rotate rigidly.

The free parameters can be constrained if one assumes that the distributions of rotation periods among Sun-like stars in open clusters of various ages result from the evolution of one initial distribution of rotation periods after dispersion of circumstellar disks. This assumption is justified by the fact that Sun-like stars in the Pleiades , M50, and M35 that have similar ages also show similar distributions of rotation periods. Rotation period measurements in the 5 Myr-old open cluster NGC 2362 by Irwin et al. (2008) provides such an initial distribution. For 0.7 - 1.1 M$_{\rm \odot}$ stars, it can be approximated to a normal distribution of rotation periods truncated below 0.3 days with a maximum around 7 days (see top left panel of Fig. 4 in Gondoin 2017). 

This method was initially applied by Gondoin (2017) using the Pleiades ($\sim$ 112 Myr), M50 ($\sim$ 130 Myr), M35 ($\sim$ 133 Myr), M37 ($\sim$ 550 Myr) open clusters and the Sun for reference comparisons. These open clusters show a broad range of rotation periods but are of a reasonable size for establishing rotation period distributions with large enough statistical significance.  Gondoin (2018) also used rotation periods measurements in NGC 6811 ($\sim$ 1 Gyr), NGC 6819 ($\sim$ 2.4 Gyr), and M67 ($\sim$ 4 Gyr). Since these old clusters show narrow distributions of rotation periods, a few stellar members with known masses and rotation periods are sufficient to constrain the rotation of solar mass stars at the corresponding ages.

Figure 1 shows the inferred evolution of the angular rotation velocity of 1.0 M$_{\rm \odot }$ stars with initial rotation periods of 1, 2, 3, 6, 10, and 20 days at an age of 5 Myrs calculated with the parametric model that best interpolate the rotation period histograms measured in the open clusters mentioned above. It shows the spin-up phase during pre-main sequence contraction followed by a spin-down near the ZAMS and during further evolution on the main sequence. The model reproduces the dispersion of rotation periods among solar-type stars in young open clusters and the narrowing of their distribution beyond an age of about 600 Myr. Remarkably, the rotation evolution model reproduces the bimodal distribution observed in the measured rotation periods histograms of open clusters aged between 30 and 600 Myr (Barnes 2003, Meibom et al. 2009, 2011).

\begin{figure*}[!ht]
	\begin{tabular}{l l}
	\centering
\includegraphics[width=0.48\linewidth, angle=0]{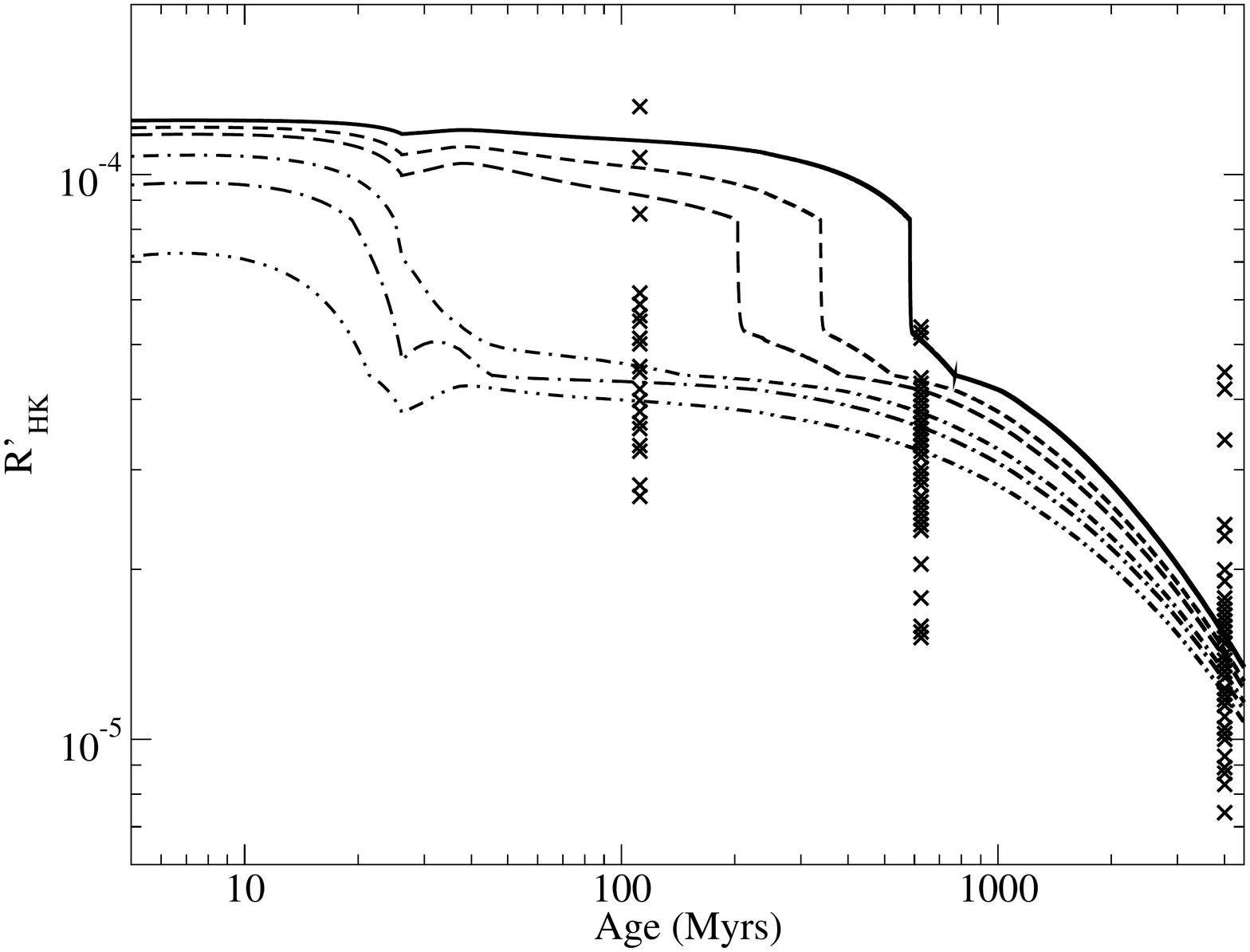} & \includegraphics[width=0.48\linewidth, angle=0]{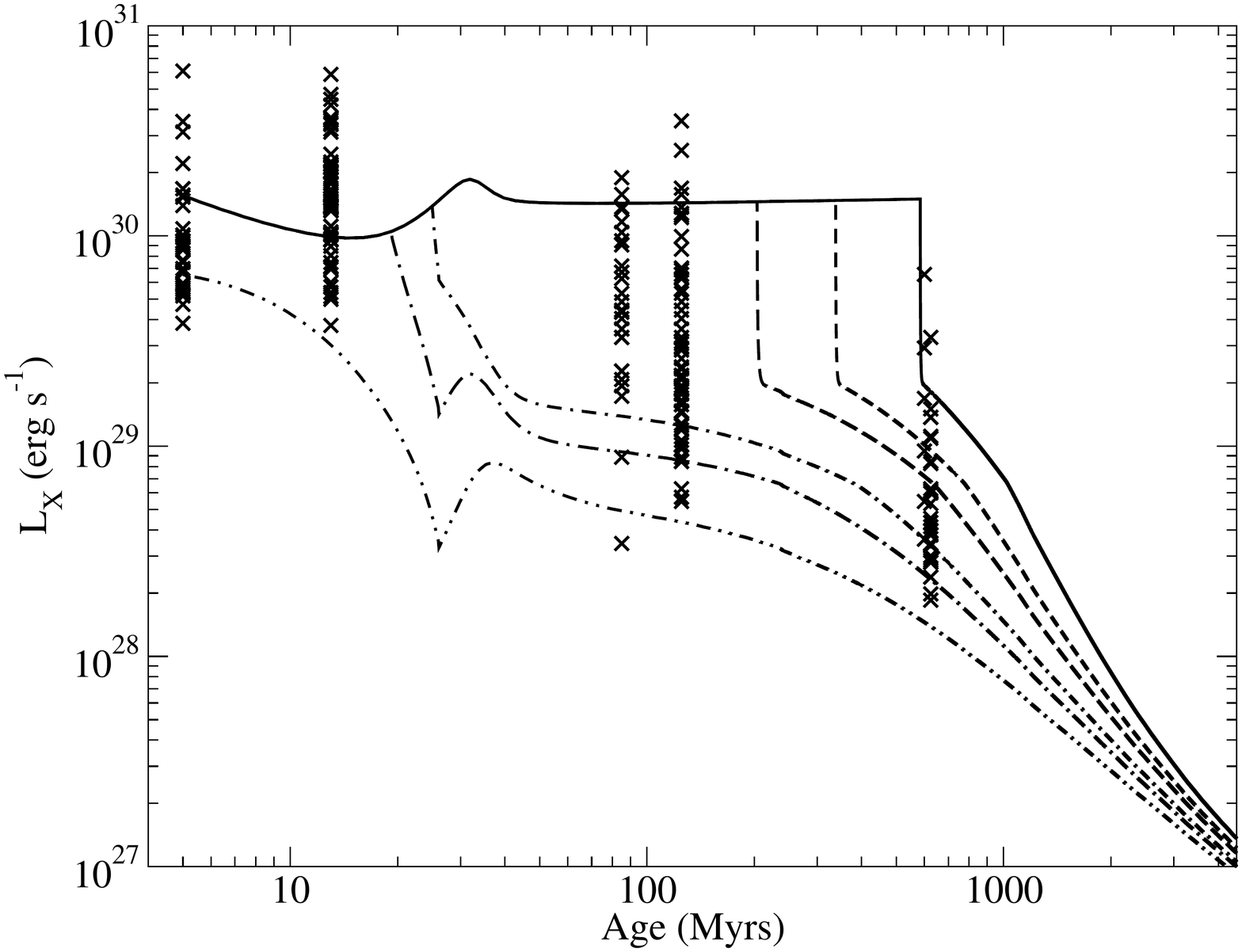} \\
\end{tabular}	
	\caption{Left: Simulated evolution of the average Ca II $R_{\rm HK}^{'}$ index (left) and average X-ray luminosity (right) of 1.0  M$_{\rm \odot }$ main-sequence stars. The different curves correspond to initial rotation periods of 1, 2, 3, 6, 10 and 20 days at an age of 5 Myrs. The curve symbols are identical to those used in the corresponding angular velocity curves of Fig. 1. The simulations of the chromospheric activity evolution are compared with measurements of the $R_{\rm HK}^{'}$ index (see Mamajek \& Hillenbrand 2008 and reference therein) on 0.9 - 1.1 M$_{\rm \odot }$ stars in the Pleiades (125 Myrs), the Hyades (625 Myrs) and M67 (4200 Myrs). The simulations of the X-ray activity evolution are compared with X-ray luminosity measurements of 0.8 to 1.2 M$_{\rm \odot}$ stars in NGC 2362 (5 Myr), h Per (13 Myr), $\alpha$ Per (85 Myr), the Pleiades (125 Myr), Praesepe (600 Myr), and the Hyades (625 Myr).}	
	\label{fig:fig_RHK}
\end{figure*}

\section{Simulated magnetic activity evolution}

\subsection{Magnetic activity on the early main sequence}

The emission reversals in the cores of the Ca II H\,\&\,K Fraunhofer lines are a well-known sign of departures from radiative equilibrium that require additional mechanisms of heating, generally termed activity (Hall 2008 and reference therein). A major contributor to chromospheric and coronal activity is the evolution and variability of magnetic fields via heating by Alfven waves or transport of mechanical energy along magnetic flux tubes into the outer atmosphere of cool stars.

Chromospheric activity has been traditionally characterised by the $R_{\rm HK}^{'}$ index, defined as the ratio of the emission in the core of the Ca II H\,\&\,K lines to the total bolometric emission of the star (Noyes et al. 1984). Mamajek \& Hillenbrand (2008) derived a unique relationship between the $R_{\rm HK}^{'}$ index of main-sequence stars and their Rossby number $Ro$ defined as the ratio between their rotation period and convective turnover time:
\begin{eqnarray}
\log R_{\rm HK}^{'} = A - B \times (Ro  - C) \nonumber \\
A, B, C = \left\{ \begin{array}{rl}
  -4.23, 1.451, 0.233 &\mbox{if Ro  $<$ 0.32} \\
  -4.522, 0.337, 0.814 &\mbox{Ro $\ge$ 0.32}.
       \end{array} \right.
\end{eqnarray}

This formula uses the convective turnover time relation as a function of $B - V$ colour of Noyes et al. (1984) based on the finding (Montesinos et al. 2001) that it produces the tightest correlation between activity and Rossby number when compared to stellar models using mixing-length theory and full turbulence spectrum treatment of convection.  

Using the best fit model of rotation evolution described in Fig. 1, I calculated the time evolution of the Rossby numbers of 1.0 M$_{\rm \odot}$ main-sequence stars assuming different initial periods of rotation after circumstellar disk dispersion and a constant convective turnover time of 14.45 days. This empirical estimate by Wright et al. (2011; see their Fig. 7) is in good agreement with that of Noyes et al. (2004). The evolution of the $R_{\rm HK}^{'}$ index was then calculated using Eq. 1. 

The results are shown in the left graph of Fig. 2 for initial rotation periods ranging from 1 to 20 days at an age of 5 Myr. The simulated curves are compared with $R_{\rm HK}^{'}$ measurements on 0.9 - 1.1 M$_{\rm \odot}$ stars in the Pleiades ($\sim$ 130 Myrs), the Hyades ($\sim$ 625 Myrs), and M67 ($\sim$ 4000 Myrs) compiled by Mamajek \& Hillenbrand (2008; see reference therein).  They reproduce the broad distribution of Ca II emission among young Sun-like stars in the Pleiades, the decay of the chromospheric emission in the intermediate age Hyades ($\sim$ 625 Myrs), and the even lower chromospheric activity of solar mass stars in the old M67 cluster. 

Coronal X-ray emission is another important magnetic activity diagnostic for cool stars. The relationship between the coronal radiative flux density and the average surface magnetic flux density has been shown to be nearly linear for solar active regions as well as for entire stars (e.g. Fisher et al. 1998; Schrijver \& Zwaan 2000) over 12 orders of magnitude in absolute magnetic flux (Pevtsov et al. 2003). 

The X-ray emission from late-type stars in open clusters exhibits two kinds of dependences on stellar rotation (e.g. Patten \& Simon 1996; Randich 2000; Feigelson et al. 2003). Fast rotators with a Rossby number smaller than $\sim$0.13 show a relatively constant X-ray to bolometric luminosity ratio at the so-called $(L_{\rm X}/L_{\rm bol})$ $\approx$ 10$^{-3}$ saturation level. Slower rotators with a larger Rossby number show a decline of their X-ray emission with decreasing rotation rate.  This behaviour can be parameterised as follows:
\begin{equation}
{L_{\rm X} \over L_{\rm bol}} \approx \left\{ \begin{array}{rl}
 R_{\rm X, sat} &\mbox{if $Ro \le Ro_{\rm crit}$} \\
 (L_{\rm X, \odot} / L_{\rm \odot})(Ro / Ro_{\rm \odot})^{\beta}  & \mbox{if $Ro >$ Ro$_{\rm crit}$}
       \end{array} \right.
\end{equation}
with $\log(L_{\rm X,\odot}/L_{\rm \odot})$ = -6.24 (Judge et al. 2003). 

In this equation, $L_{\rm X}$ is the stellar X-ray luminosity,  $L_{\rm bol}$ the bolometric luminosity and $Ro_{\rm crit}$ the Rossby number value below which the saturation of X-ray emission occurs.  R$_{\rm X, sat}$ = 0.74 $\times$ 10$^{-3}$ is the saturation level of the X-ray to bolometric luminosity ratio measured in the $\it{ROSAT}$ 0.1 - 2.4 keV energy band. Pizzolato et al. (2003) found a power index $\beta$ = -2 while Wright et al. (2011) argued that a value $\beta$ = -2.70 $\pm$ 0.13 provides a better fit to the Sun's X-ray luminosity. 

By combining Eq. 2 with the best fit model of rotation evolution assuming a convective turnover time of 14.45 days,  I calculated the X-ray luminosity evolution of 1.0 M$_{\rm \odot }$ main-sequence stars with initial rotation periods ranging from 1 to 20 days at an age of 5 Myr. The results are plotted in the right graph of Fig. 2. They are compared with X-ray luminosity measurements of 0.9 - 1.1 M$_{\rm \odot}$ stars in $\alpha$ Per (85 Myrs), the Pleiades (125 Myrs), Praesepe (600 Myrs), and the Hyades (625 Myrs) that have been compiled by Wright et al. (2011; see reference therein). 

Figure 2 shows that the simulated evolution of X-ray luminosity on solar-mass stars is similar to that of their $R_{\rm HK}^{'}$ index. Between the ages of  $\sim$50 and $\sim$600 Myr, both magnetic activity indicators show a large dispersion. Fast rotators on the early main-sequence exhibit a very similar and large magnetic activity level at $R_{\rm HK}^{'}$ $\approx$  10$^{-4}$ and $L_{\rm X}/L_{\rm bol}$ $\approx$ 10$^{-3}$.  In contrast, the slow rotators on the early main-sequence show moderate activity levels around $log (R_{\rm HK}^{'})$ $\approx$  -4.4 and $L_{\rm X}$ $\approx$ 10$^{29}$ erg s$^{-1}$. Beyond an age of about 600 Myr, the magnetic activity levels of both populations decay converging towards similar values by the age of the Sun.

\begin{figure*}[!ht]
\begin{tabular}{c c c}
	\centering	
 \includegraphics[width=0.31\linewidth, angle=0]{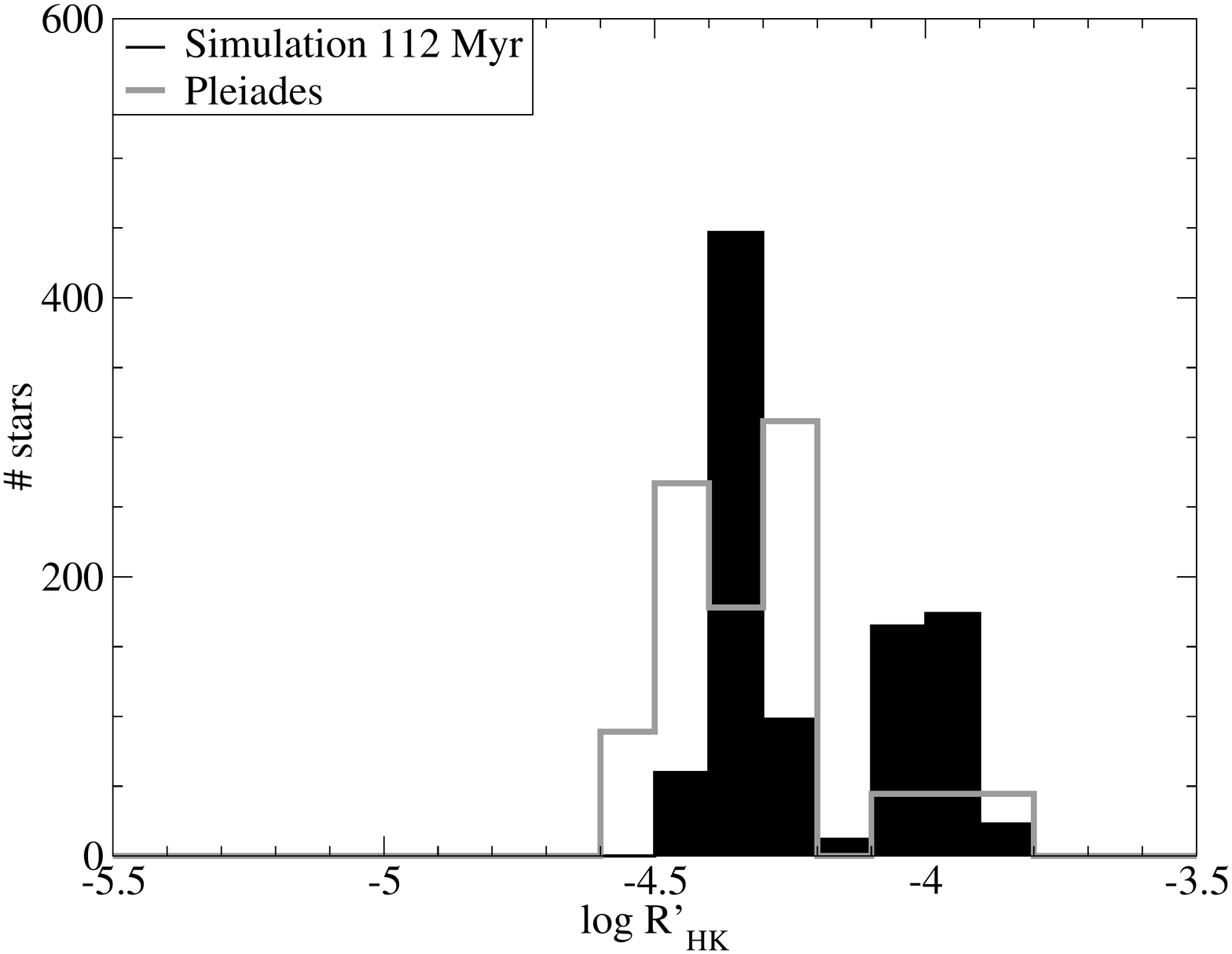}  & \includegraphics[width=0.31\linewidth, angle=0]{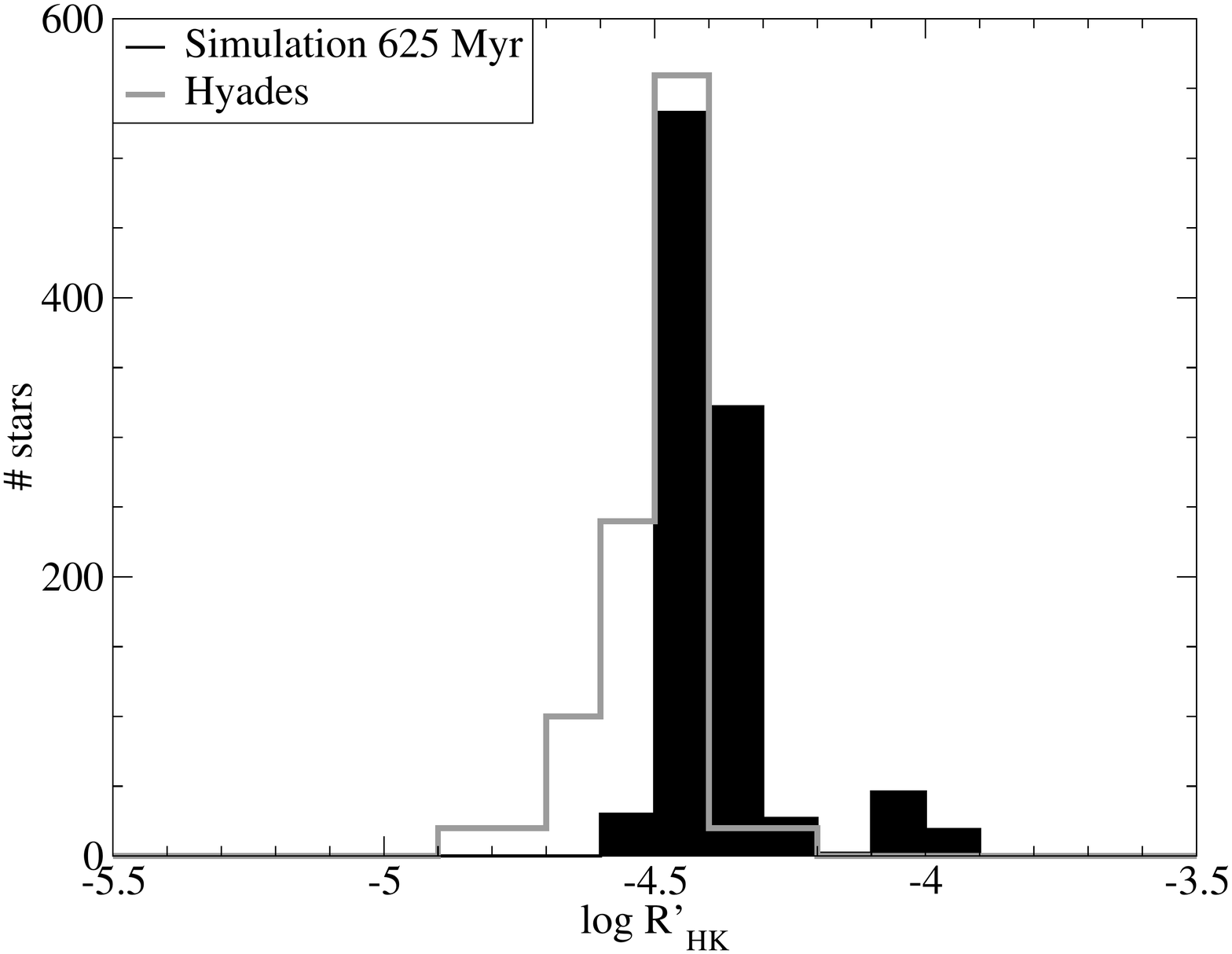} & \includegraphics[width=0.31\linewidth, angle=0]{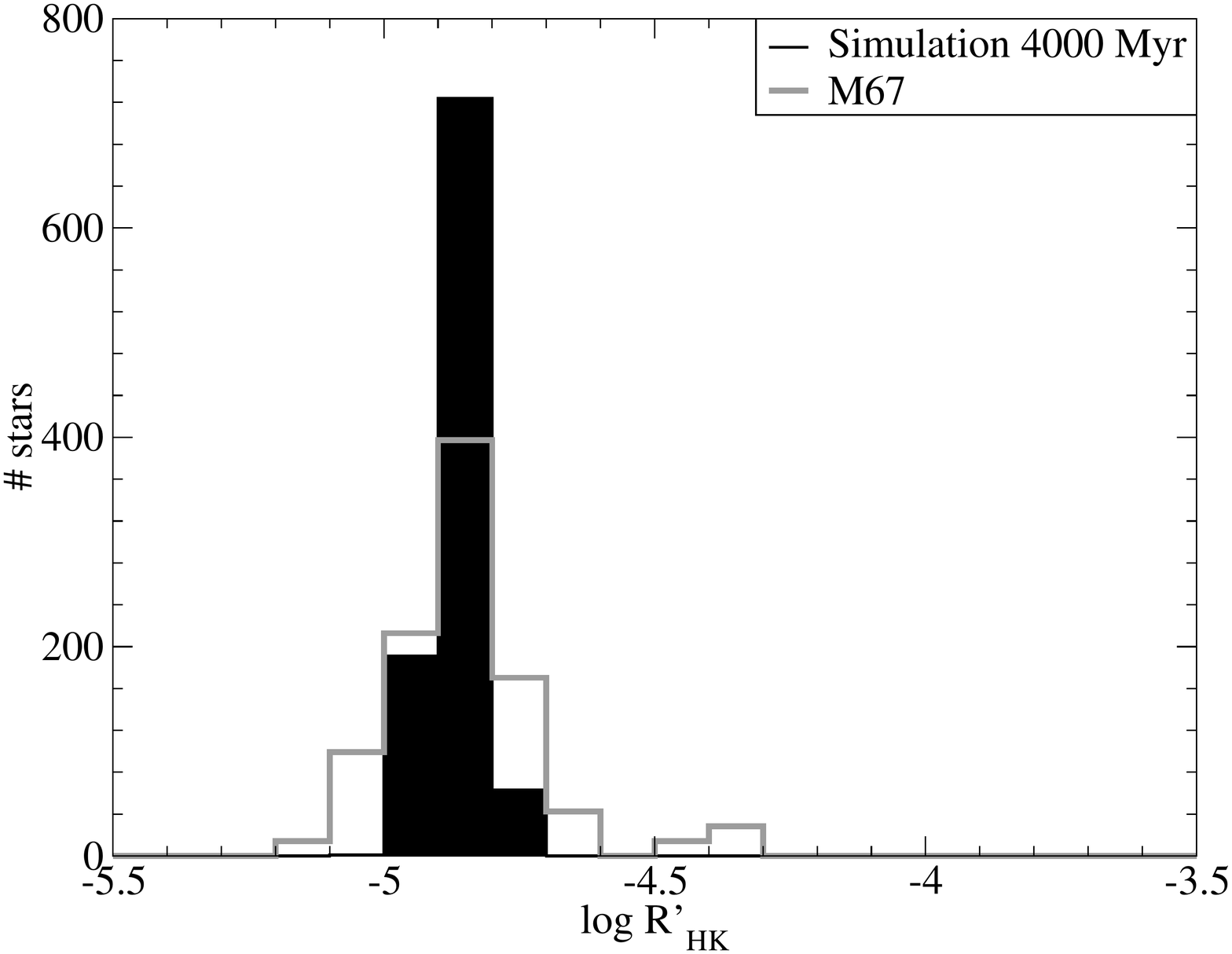} \\   
 \end{tabular}
	\caption{Simulated $R_{\rm HK}^{'}$  index distributions of solar mass stars in open clusters with ages of 112, 625, and 4000 Myr. The results are compared with measured $R_{\rm HK}^{'}$ histograms of  0.9 -1.1 M$_{\rm \odot}$ stars in the Pleiades, the Hyades, and M67 scaled to the 979 stars used in the simulation. The short-term variability of the magnetic activity is described by a normal distribution of the short-term CaII chromospheric emission with a variance of 0.327 $\times <R_{\rm HK}^{'}>^{1.15}$ around the average chromospheric emission $<R_{\rm HK}^{'}>$ derived from a parametric model of rotation evolution for solar mass stars.}
	\label{fig:fig_meas_rot_distrib}
\end{figure*}

\subsection{Magnetic activity on the pre-main sequence}

Since the rotation-activity relationships (see Eq. 1 and 2) have been established empirically considering only the colour or mass dependence of the convective turnover time but ignoring any stellar age dependent effect, they are a priori only applicable to main-sequence (MS) stars.  However, pre-main sequence (PMS) stars are also magnetically active, producing strong magnetic fields and intense coronal emission (e.g. Preibisch \& Feigelson 2005, Gondoin 2006, Donati et al. 2012, Vidotto et al. 2014). 

In the past decade, photometric and X-ray surveys of young open clusters have produced rotation period and X-ray luminosity measurements on pre-main sequence stars of different ages.  Some studies have investigated the properties of coronal activity among PMS stars within cluster younger than $\sim$3 Myr. A large fraction of these very young stars is accreting material from circumstellar disks producing poor correlations between their X-ray and bolometric luminosities  or stellar masses and a large scatter (e.g., Stassun et al. 2004; Briggs et al. 2007). Nevertheless, the X-ray luminosity of non-accreting T Tauri stars has been found to show well-defined correlation with their bolometric luminosity similarly to rapidly rotating MS stars (Preibisch et al. 2005). 

A few studies have also investigated the relationship between X-ray activity and rotation among PMS stars older than $\sim$3 Myr that have cleared most of their circumstellar disks but have not yet reached the ZAMS.  In a search for correlations between X-ray emission and rotation among PMS stars in the h Per open cluster that is 13 Myr old, Argiroffi et al. (2016) conclude that solar-like members of this cluster show different activity regimes, analogous to those observed among MS stars. Samples of 38 Myr old stars in NGC 2547 (Jeffries et al. 2006; Irwin et al. 2008) also follow the X-ray to bolometric luminosity vs Rossby number relationship observed by Pizzolato et al. (2003) among a range of solar-type stars from the field, the Pleiades and other open clusters. 

These results suggest that PMS and MS stars follow the same rotation-activity relationships. I thus calculated the evolution of the chromospheric activity index and X-ray to bolometric luminosity ratio on the pre-main sequence by combining the best fit models of rotation evolution (see Fig. 1) with the  rotation-activity relationships given by Eq. 1 and 2. The convective turnover time that had been assumed to be constant for MS stars was replaced by an age dependent function emulating the convective turnover time evolution of a solar mass star calculated by Landin et al. (2010).  This convective turnover time is age dependent on the pre-main sequence. It converges towards an almost constant value on the main sequence close to the 14.45 days derived empirically by Wright et al. (2011) for a solar mass star.  

The decay of the convective turnover time during PMS evolution has a significant effect on the Rossby number. Since both the rotation period and the convective turnover time decrease as PMS stars contract towards the main sequence, the Rossby number of a solar mass star remains small till an age of $\sim$30 Myr. During the contraction phase, only the small fraction of stars with initially long periods of rotation (P$_{\rm 0}$ $\gtrsim$ 20 d at 5 Myr) have Rossby numbers greater than the critical value Ro $\approx$ 0.13 above which activity is correlated with rotation. Applying the rotation-activity relationships to PMS stars thus results in most of these stars operating in a rotation-independent regime of magnetic activity. 

This behaviour is illustrated in Fig. 2. Most stars younger than about 30 Myrs operate at high levels of chromospheric and coronal activity with $R_{\rm HK}^{'} \approx$ 10$^{-4}$ and  $L_{\rm X}/L_{\rm bol} \approx$ 10$^{-3}$, respectively. These simulation results are consistent with measured X-ray luminosities of 0.9 - 1.1 M$_{\rm \odot}$ stars in NGC 2362 (Damiani et al. 2006) and h Per (Argiroffi et al. 2016) that are 5 and 13 Myr old, respectively.  

\section{Simulated vs measured activity indices}

\subsection{The short-term variability of magnetic activity}

Figure 2 describes the effect of stellar rotation on the long-term evolution of chromospheric and coronal emission on solar mass stars. While stellar rotation evolves slowly with time, magnetic activity varies on much shorter timescales. The physical mechanisms producing such variability include changes in the filling factor of active regions, growth and decay of individual emitting regions, and activity cycles. 

By exploring the variation in the photometric bands centred in the Ca II H and K lines on timescales of years, it has been found (Baliunas et al. 1995) that young Sun-like stars tend to vary irregularly, rather than in a smooth cycle like that of the Sun. In contrast, older more slowly rotating stars have weaker average activity levels, often with regular, cyclic variation superimposed. The timescales of these cycles range from a few years to more than a decade.

Using chromospheric time series of stars with spectral type F5 to K7 observed between 1984 and 1995,  Radick et al. (1998) explored the relation between chromospheric variability and average activity level. These authors found that chromospheric variations can be fairly well related by power laws to average chromospheric activity levels for their entire sample of stars that spanned a range of chromospheric activity indices between -5.05 and -4.15. An estimate of the cyclic chromospheric RMS variation (see Fig. 5 in Radick et al. 1998) is given by:
\begin{equation}
RMS(<R_{\rm HK}^{'}>) = 0.327 \times <R_{\rm HK}^{'}>^{1.15}
\end{equation}

The $\it{Einstein}$ IPC and $\it{ROSAT}$  PSPC, taken roughly a decade apart, have provided valuable information on the variability of X-ray stellar activity in the Pleiades and the Hyades. Temporal analyses of Pleiades X-ray data, using single $\it{ROSAT}$ observations, have been made by Schmitt et al. (1993), Stauffer et al. (1994), and Micela et al. (1996). Micela et al. (1996) found that most of the Pleiades stars show variability within a factor two-three, both on six months and, using Einstein observations, on ten years timescales. A few stars with evidence of larger variations show large flares on shorter time scales. Gagne et al. (1995) found that on one year timescales, approximately 25\% of the late-type Pleiades stars are variable by more than a factor of two, while they found only a marginal evidence for increased variability on the ten-year timescale. Regarding the Hyades, Stern et al. (1995) found that more than 90\% of the stars common to the $\it{Einstein}$ IPC and $\it{ROSAT}$ PSPC surveys show variability of less than a factor of two. 

Based on these observations, I estimated that stellar X-ray luminosities vary on short timescale by a factor of three in the Pleiades at a 99\% confidence level and by a factor of two in the Hyades at a 95\% confidence level.  Assuming a normal distribution of the instantaneous X-ray luminosity around the average X-ray activity level $<L_{\rm X}>$ at a given age, the standard deviation of the coronal X-ray luminosity at these ages would thus be:
\begin{equation}
RMS(<L_{\rm X}>) = <L_{\rm X}>/6
\end{equation}

\begin{figure*}[!ht]
\begin{tabular}{c c c}
	\centering
 \includegraphics[width=0.31\linewidth, angle=0]{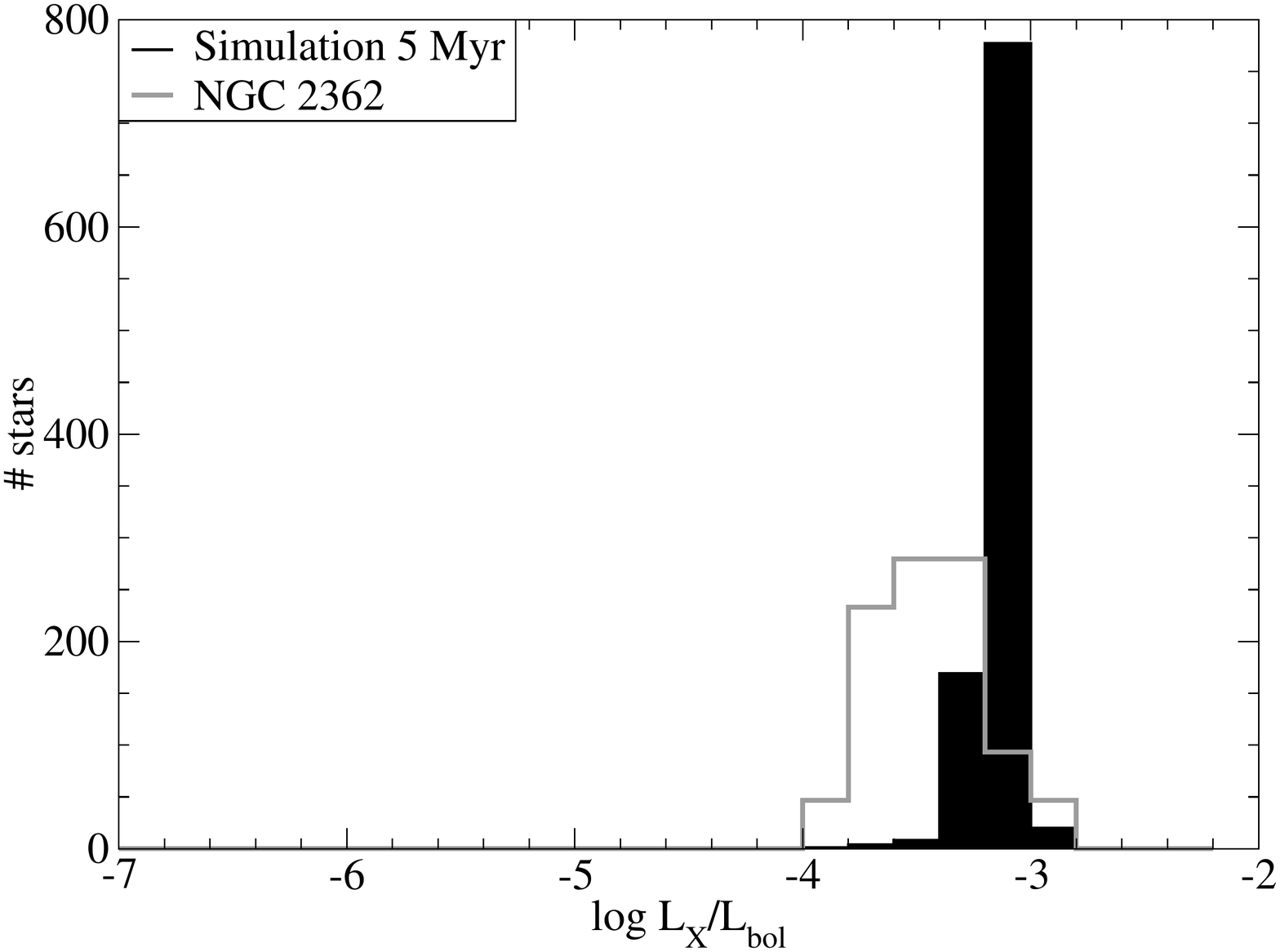} & \includegraphics[width=0.31\linewidth, angle=0]{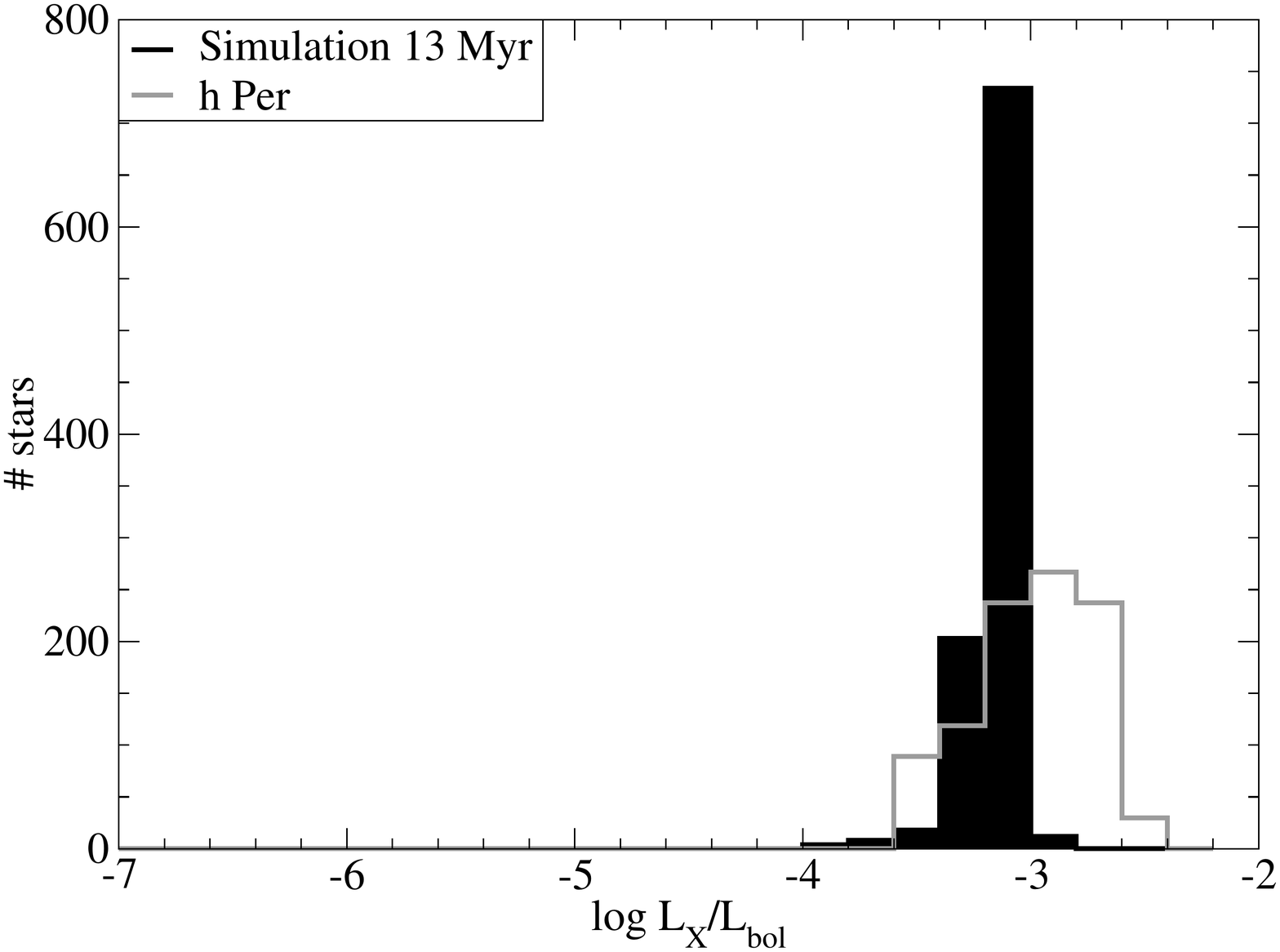} & \includegraphics[width=0.31\linewidth, angle=0]{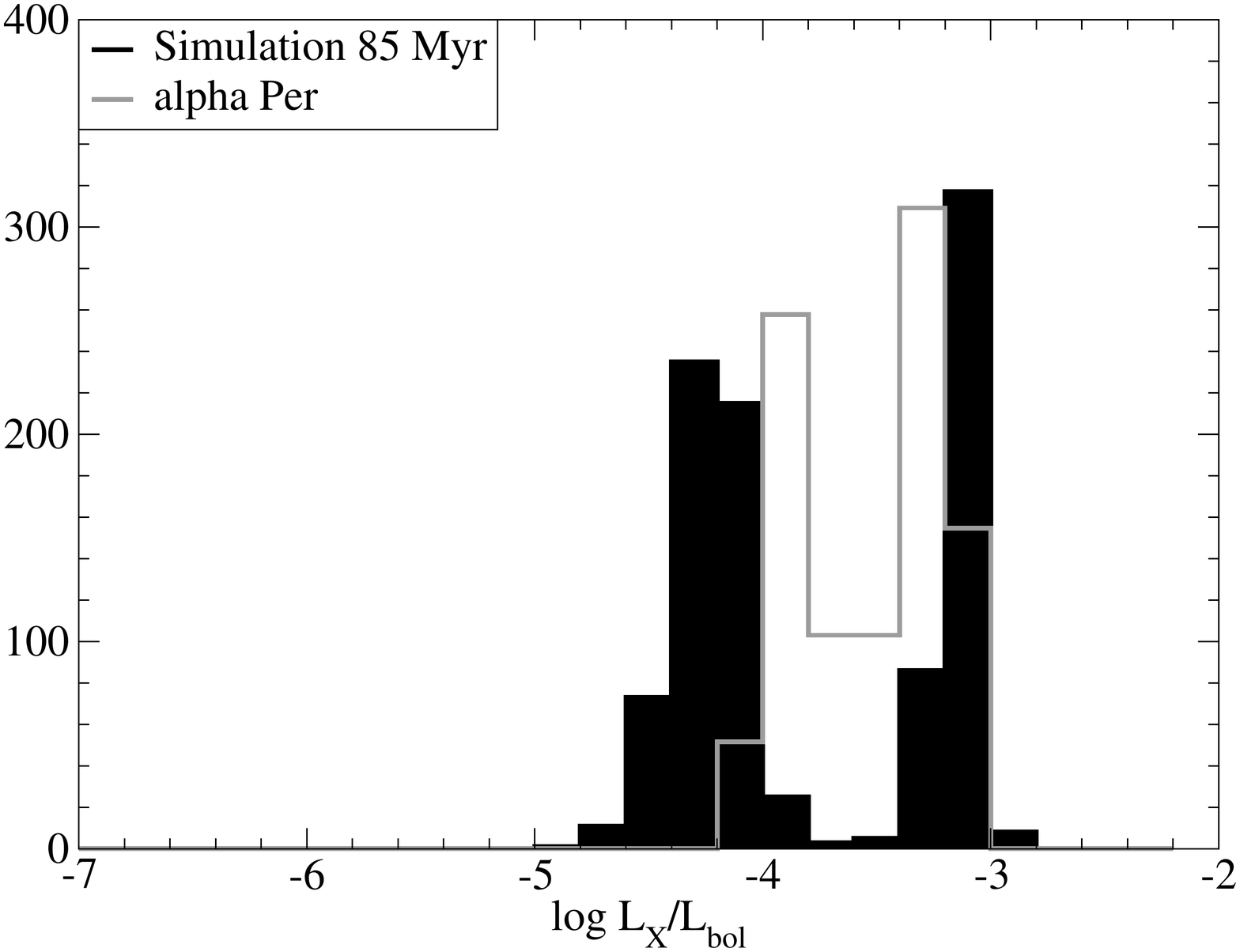} \\ 	
 \includegraphics[width=0.31\linewidth, angle=0]{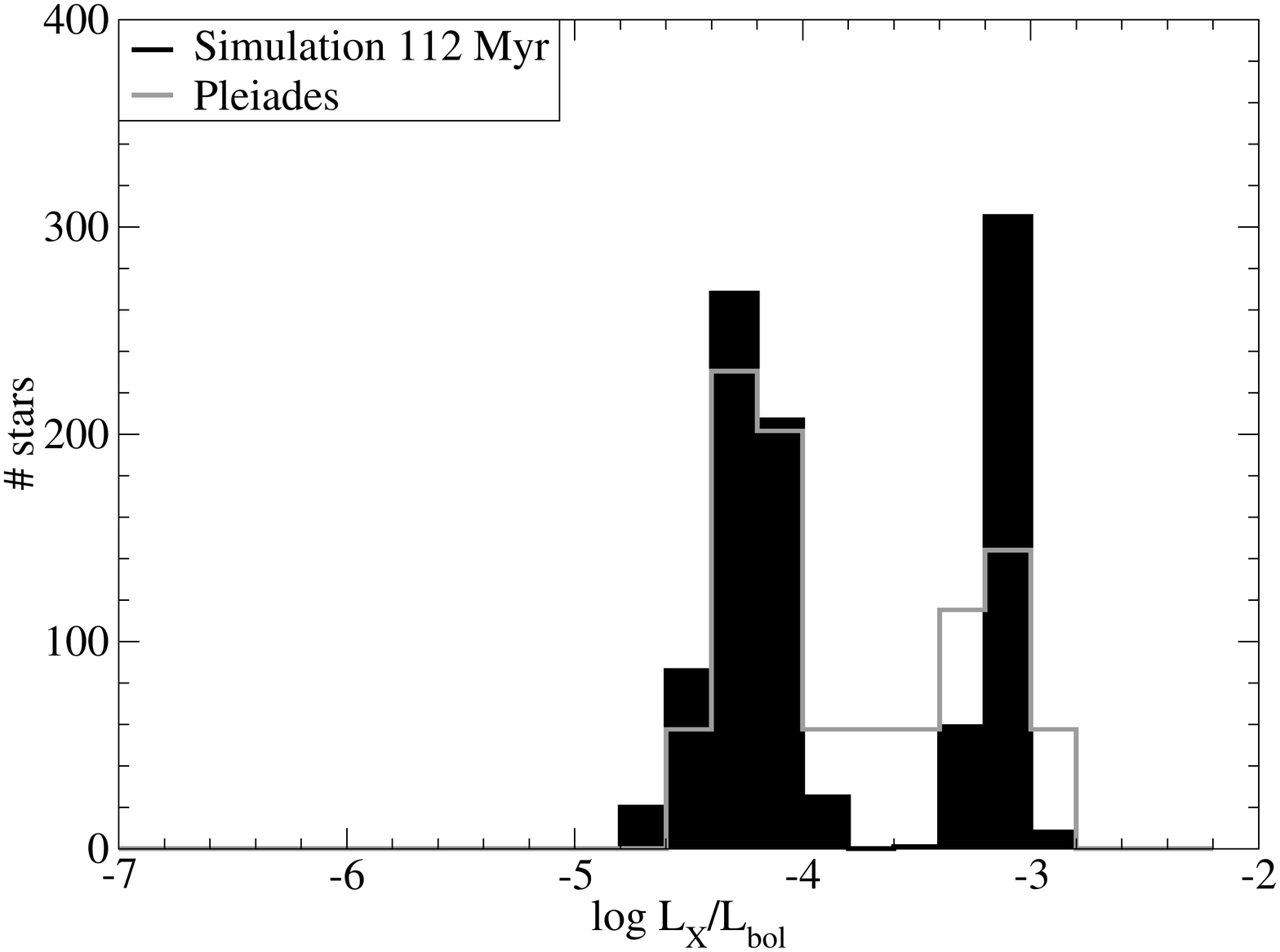}  & \includegraphics[width=0.31\linewidth, angle=0]{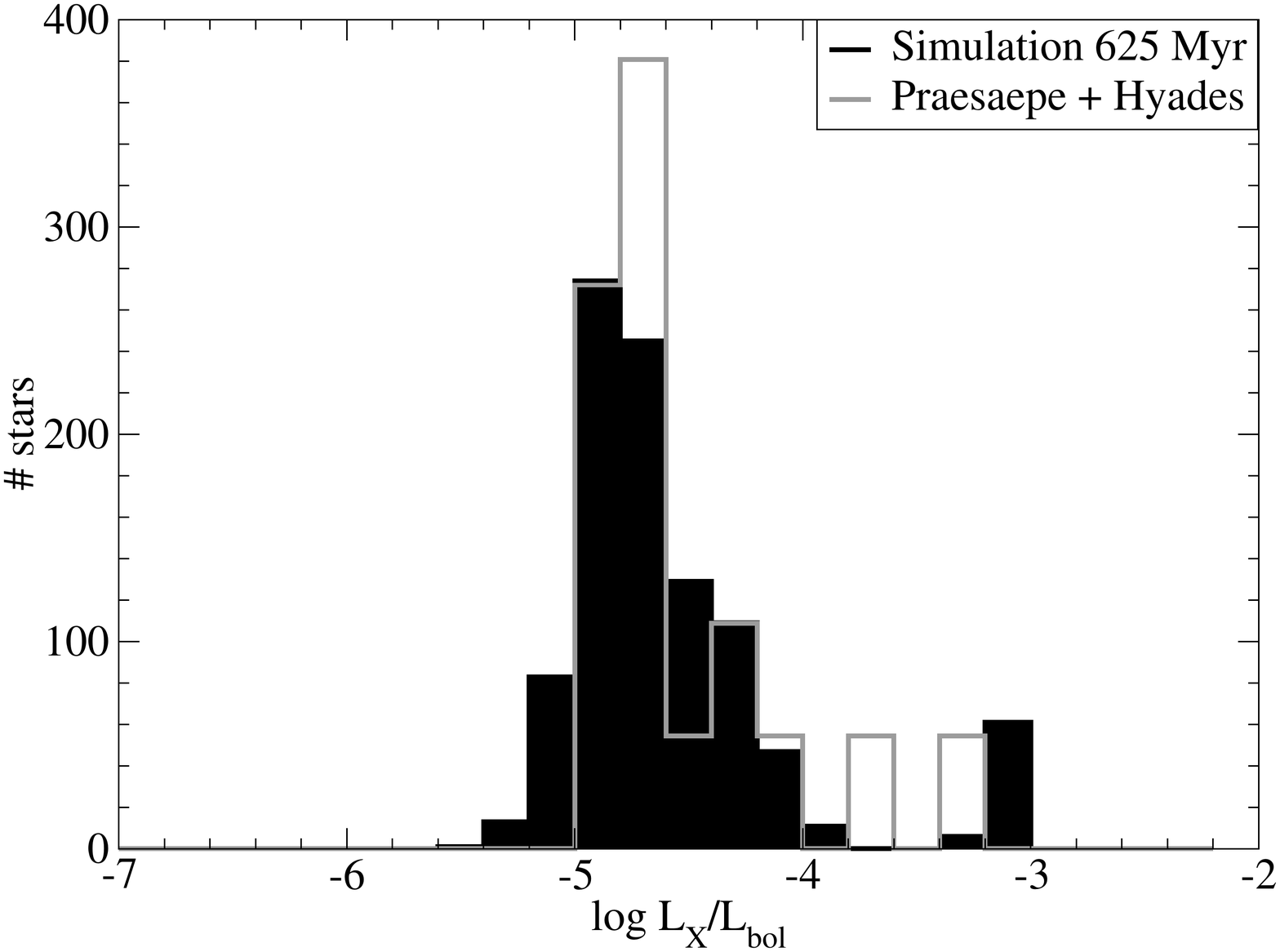} & \includegraphics[width=0.31\linewidth, angle=0]{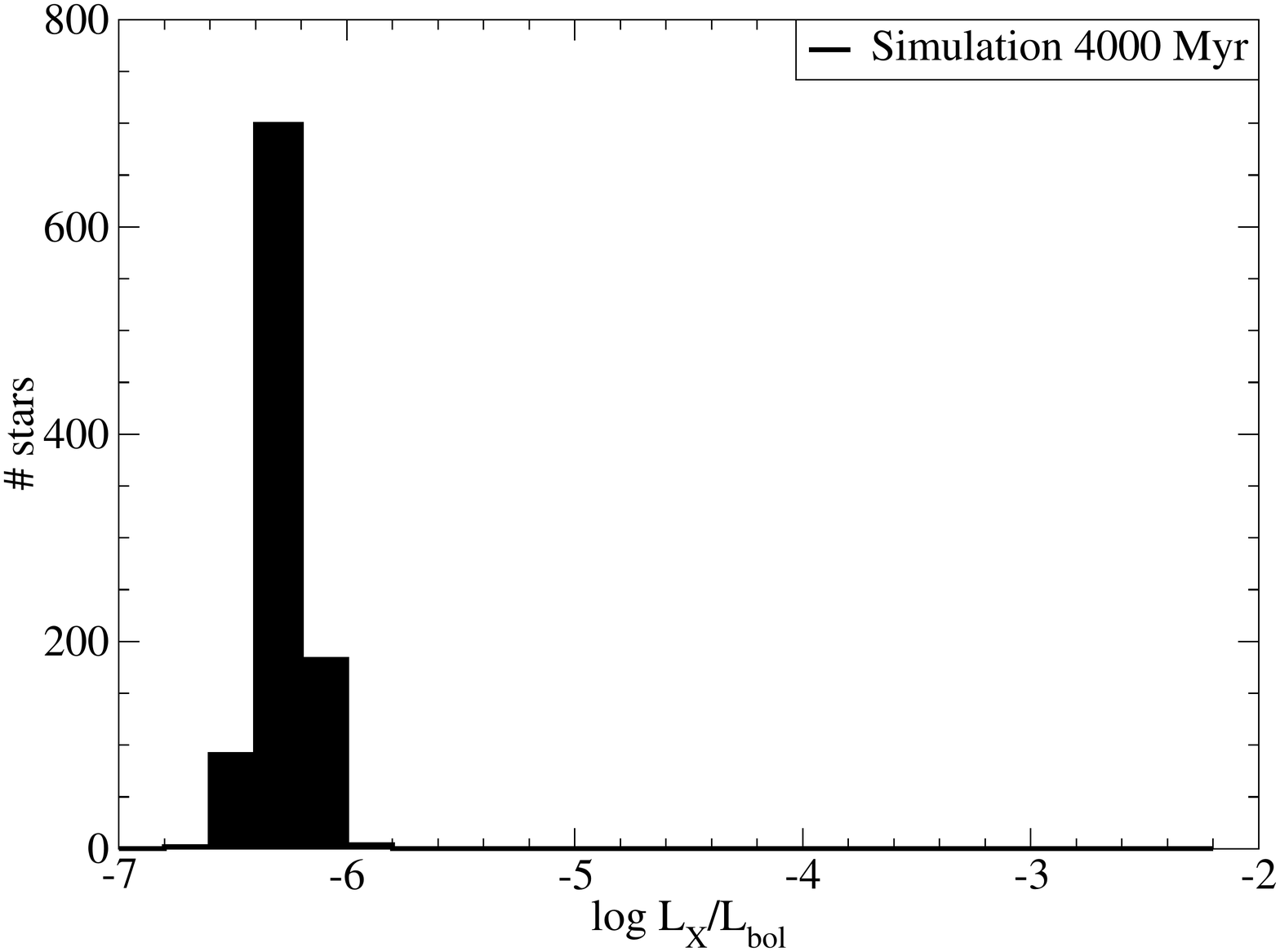} \\   
 \end{tabular}
	\caption{Simulated $L_{\rm X}/L_{\rm bol}$ ratio distributions of solar mass stars in open clusters with ages of 5, 13, 85, 112, 625, and 4000 Myr. The results are compared with measured $L_{\rm X}/L_{\rm bol}$ histograms of  0.9 -1.1 M$_{\rm \odot}$ stars in NGC 2362, h Per, $\alpha$ Per, the Pleiades, Praesaepe and the Hyades scaled to the 979 stars used in the simulation. The short-term variability of the magnetic activity is described by a normal distribution of the short-term X-ray emission with a variance of $<L_{\rm X}>$/6 around the average X-ray luminosity $<L_{\rm X}>$ derived from a parametric model of rotation evolution for solar mass stars.}
	\label{fig:fig_meas_rot_distrib}
\end{figure*}

\subsection{Activity indices distributions in open clusters}

The distribution of rotation periods among Sun-like stars in open clusters older than $\sim$5 Myr can be estimated by applying a best fit model of rotation evolution to an initial distribution of stellar rotation periods after circumstellar disk dispersion. At 5 Myr, this initial distribution can be approximated by a normal distribution truncated below 0.3 days with a maximum around seven days (Gondoin 2017).  The distribution of activity indices among Sun-like stars in any open cluster older than 5 Myr can thus be inferred by applying the rotation-activity relationships to these calculated distributions of rotation periods. 

I  applied this method to calculate the $R_{\rm HK}^{'}$  indices and X-ray to bolometric luminosity ratios of solar mass stars in open clusters with ages between 5 Myr and 4 Gyr. To simulate the short-term variability of the magnetic activity, the results were convoluted with normal distribution of activity indices using the standard deviations provided by Eqs. 3 and 4. The obtained histograms of  $R_{\rm HK}^{'}$ and  $L_{\rm X}/L_{\rm bol}$ indices are plotted in Figs. 3 and 4, respectively.

Simulations of $R_{\rm HK}^{'}$ index distributions in open clusters show that solar mass stars on the pre-main sequence are distributed around a maximum of chromospheric activity at $\log (R_{\rm HK}^{'}) \approx$ -4.0. During the early main-sequence evolution, this single-peak distribution evolves into a bimodal distribution around $\log (R_{\rm HK}^{'}) \approx$ -4.4 and $\log (R_{\rm HK}^{'}) \approx$ -4.0. By the age of the Hyades, the high activity peak of the bimodal distribution has almost entirely disappeared. The remaining peak then drifts from moderate activity levels around $\log (R_{\rm HK}^{'}) \approx$ -4.5 at an age of 1 Gyr to low activity levels around $\log (R_{\rm HK}^{'}) \approx$ -4.9 at 4 Gyr. 

Figure 3 compares the simulated distributions with measured histograms of chromospheric activity indices among 0.9 - 1.1 M$_{\rm \odot}$ stars in the Pleiades (Duncan et al. 1991; Soderblom et al. 1993; White et al. 2007), the Hyades (Duncan et al. 1991; Paulson et al. 2002; White et al. 2007), and M67 (Giampapa et al. 2006) compiled by Mamajek and Hillenbrand (2008). The total number of stars in the measured histograms of the Pleiades (22 stars), the Hyades (49 stars), and M67 (69 stars) are scaled to the number of Monte-Carlo simulations (979). Despite the limited number of measurements, Fig. 3 suggests that solar mass stars in the Pleiades that are 112 Myr old indeed group into two distinct activity levels around $\log (R_{\rm HK}^{'}) \approx$ -4.5 to -4.2 and $\log (R_{\rm HK}^{'}) \approx$ -4.1 to -3.8. In contrast, the distribution of measured activity indices in the Hyades that are 625 Myr old shows a single peak around $\log (R_{\rm HK}^{'}) \approx$ -4.5. The measured $R_{\rm HK}^{'}$ histogram of M67 that is $\sim$4 Gyr also shows a single peak but at a lower chromospheric activity level -5.1 $< \log (R_{\rm HK}^{'}) <$ -4.6. The measurements thus look consistent with the simulated distributions. 

The simulated distributions of X-ray to bolometric luminosity ratios among solar mass stars in open clusters are similar to their calculated  $R_{\rm HK}^{'}$ index distributions. Indeed, the simulations indicate that PMS stars emit X-rays at the saturation level around $L_{\rm X}/L_{\rm bol} \approx$ 10$^{-3}$. On the early main-sequence, this single peak distribution evolve into a bimodal distribution around $\log(L_{\rm X}/L_{\rm bol}) \approx$ -3.0 and $\log(L_{\rm X}/L_{\rm bol}) \approx$  -4.1. By the age of Praesaepe and the Hyades, the high activity peak of the bimodal distribution has almost entirely disappeared. The remaining single-peak distribution then drift from moderate activity levels around $\log(L_{\rm X}/L_{\rm bol}) \approx$ -5.0 at an age of 1 Gyr to low activity levels around $\log(L_{\rm X}/L_{\rm bol}) \approx$ -6.1 at 4 Gyr.

Figure 4 compares simulated histograms with measured $L_{\rm X}/L_{\rm bol}$ histograms of  0.9 -1.1 M$_{\rm \odot}$ stars in NGC 2362,  h Per, $\alpha$ Per, the Pleiades, the Hyades and Praesaepe. It shows that PMS stars in NGC 2362 and h Per have X-ray to bolometric luminosity ratios close to the so-called saturation level of X-ray emission at $L_{\rm X}/L_{\rm bol} \approx$ 10$^{-3}$. Two peaks are observed in the measured distribution of X-ray luminosities in $\alpha$ Per and the Pleiades. The measured distribution of X-ray to bolometric luminosity ratio in Praesaepe and the Hyades shows a strong  peak at $L_{\rm X}/L_{\rm bol} \approx$ 1 - 2.5$\times$10$^{-5}$ with very few stars having higher X-ray luminosities. These measurements are consistent with the simulated distributions of X-ray to bolometric luminosity ratio in those open clusters. They provide additional evidences of the onset and disappearance of a bimodal distribution of activity levels in intermediate-age open clusters. 

\section{Summary}

A scenario of magnetic activity evolution on Sun-like stars is inferred combining a parametric model of their rotation evolution with empirical rotation-activity relationships. The best fit parameters of the rotation model are obtained by comparing measured and simulated distributions of rotation periods in open clusters of various ages assuming that they result from the evolution of a same initial distribution. The study uses rotation periods measurements of  0.9 -1.1 M$_{\rm \odot}$ stars in NGC 2362 ($\sim$ 5 Myr), the Pleiades ($\sim$ 112 Myrs), M50 ($\sim$ 130 Myr), M35 ($\sim$ 133 Myr), M37 ($\sim$ 550 Myr), NGC 6811 ($\sim$ 1.0 Gyr), NGC 6819 ($\sim$ 2.4 Gyr), and M67 ($\sim$ 4.0 Gyr). It assumes that PMS and MS stars follow the same relationships between Rossby number and chromospheric or coronal emission.  An age-dependent estimate of the local convective turnover time is used in the calculation of the Rossby number.

The inferred scenario of magnetic activity evolution (see Fig. 2)  reproduces the high levels of chromospheric and coronal emission ($R_{\rm HK}^{'} \approx$ 10$^{-4}$ and  $L_{\rm X} \approx$ 10$^{30}$ erg s$^{-1}$, respectively)  observed on solar mass stars during their early pre-main sequence evolution.  It indicates that, at the end of the PMS contraction phase around the age of $\sim$ 30 Myr, the slowest rotating stars experience a rapid transition to moderate activity levels around $R_{\rm HK}^{'} \approx$ 4 $\times$ 10$^{-5 }$ and $L_{\rm X} \approx$ 10$^{29}$ erg s$^{-1}$. This brief episode of rapid decay of the magnetic activity occurs later on more rapidly rotating stars, up to an age of $\sim$600 Myrs for the fastest rotators. After this sharp transition, the average chromospheric and coronal activity indices of all stars decrease steadily converging towards similar values ($R_{\rm HK}^{'} \approx$ 10$^{-5}$ and  $L_{\rm X} \approx$ 10$^{27}$ erg s$^{-1}$) by the age of the Sun. 

Provided that the short-term variability of magnetic activity is taken into account, simulated distributions of activity indices at 5, 13, 85, 112, 625, and 4000 Myrs are similar to measured histograms in  NGC 2362, h Per,  $\alpha$ Per, the Pleiades, the Hyades, and M67 ($\sim$ 4 Gyr) respectively.  In particular, both simulation and measurements indicate that clusters with age between 30 and 600 Myr exhibit a bimodal distribution of chromospheric activity and coronal emission. The currently available measurements of activity indices in open clusters are thus consistent with the rotation evolution model of Sun-like stars proposed by Gondoin (2017) and with the corresponding scenario of their magnetic activity evolution (Gondoin 2018). This corroborates the hypotheses that (i) similar rotation-activity relationships hold for MS and PMS stars, (ii) a brief episode of large angular momentum loss occurs in the early evolution of Sun-like stars and (iii) the distributions of rotation rates among Sun-like stars in open clusters depend mainly on their age, that is, they derive from similar initial distributions of rotation periods after dispersion of their circumstellar disks. 

\section*{Acknowledgments}
{I am grateful to the organising committee of the Cool Stars 20 workshop for allowing me to present this work.}


\end{document}